\documentclass[a4paper,10pt]{article}
\usepackage{epsfig}
\usepackage{graphicx} % Di Domenico
\usepackage{amsmath} %LiGioi

\newcommand {\Bqz}{{\overline{B}}{}^0}\newcommand {\Bz}{B^0} 
\newcommand {\Kz}{K^0} \newcommand {\Kqz}{{\overline K}{}^0}
\newcommand {\Abar}{{\overline A}}

\newcommand {\Begeq}{\begin{equation}} \newcommand {\Endeq}{\end{equation}}
\newcommand {\bEa}{\begin{eqnarray}} \newcommand {\eEa}{\end{eqnarray}}
\newcommand {\imla}{{\rm Im}\lambda} \newcommand {\imlaT}{{\rm Im}{\tilde\lambda}}

\newcommand {\T}{$T$~} \newcommand {\CP}{$CP$~} \newcommand {\babar}{$BABAR$~} \newcommand {\B}{$B$~}
\newcommand {\Vub}{|V_{ub}|} \newcommand {\KS}{K^0_S} \newcommand {\KL}{K^0_L}
\newcommand {\BFs}{$B$~Factories} \newcommand {\FourS}{$\Upsilon(4S)$} \newcommand {\belle}{Belle~}
\newcommand {\invab}{\,ab$^{-1}$} \newcommand {\CPT}{$CPT$~} \newcommand {\D}{$D$~}

\newcommand{\kn}{K^0~} \newcommand{\knb}{{\overline{K}{}^0}~} \newcommand{\kpp}{K_+ ~} \newcommand{\knn}{K_-}
\newcommand{\knnp}{{\widetilde K_-}~} \newcommand{\kln}{K_L~} \newcommand{\ksn}{K_S~} \newcommand{\kppp}{{\widetilde K_+}~}

\begin{document}

\begin{flushright} MITP/13-089\\ 12 July 2013\end{flushright}
\noindent {\bf \large Conclusions of the MITP Workshop on\\[1mm] T Violation and CPT Tests in 
Neutral-Meson Systems}\\[4mm]
\noindent
{\bf K.~R.~Schubert$^{1,2}$, L.~Li~Gioi$^3$, A.~J.~Bevan$^4$ and A.~Di~Domenico$^5$}\\[4mm]
\noindent
Date of the Workshop: 15 - 16 April 2013\\[4mm]
Participants: J.~Bernabeu$^6$, A.~J.~Bevan$^4$, G.~D'Ambrosio$^7$, A.~Denig$^2$, A.~Di~Domenico$^5$,\\
H.-J.~Gerber$^8$, W.~Gradl$^2$, M.~Heck$^9$, T.~Hurth$^2$, J.~S.~Lange$^{10}$, L.~Li~Gioi$^3$, M.~Neubert$^2$,\\ 
T.~Ruf$^{11}$, K.~R.~Schubert$^{1,2}$ and P.~Villanueva-Perez$^6$\\[4mm]
The two-day workshop took place in the Institut f\"ur Kernphysik, Universit\"at Mainz in 
a very lively and friendly atmosphere, and the participants thank MITP very much for providing 
the frame for our presentations and discussions. Half of the time was used for 
discussions, and in fact the workshop continued by a number of e-mail exchanges 
until summer 2013.
This summary does not cover all contributions, but they are
available on the Indigo page of the workshop \cite{indigo}. The four parts of the summary
are:
\begin{enumerate}\itemsep -2pt
\item{T Violation in Decays of Neutral B Mesons, K.~R.~Schubert}
\item{T and CPT studies in $\Bz\Bqz$ transitions with Belle, L.~Li Gioi}
\item{Future Measurements of T violation in B and D decays, A.~J.~Bevan}
\item{Direct tests of T and CPT symmetries in the entangled neutral
Kaon system at a $\Phi$ factory, A.~Di Domenico}
\end{enumerate}
The first part covers the central point of the workshop, the interpretation of CP and
T violation in the interplay of $\Bz\Bqz$ transitions and decays $B\to J/\psi K$. It
also covers the continued and very helpful discussions with H.-J.~Gerber, T.~Ruf, 
F.~Martinez-Vidal, P.~Villanueva-Perez and A.~Di Domenico. Parts 2 to 4 cover the prospects of future
experiments with K, D and B mesons.\\

The most sensitive tests of CPT symmetry remain the Bell-Steinberger analyses
of the $\Kz\Kqz$ system using unitarity which connects the CP-symmetry properties of all observed $K_S$ and $K_L$ decay modes with the CPT- and T-sensitive overlap $\langle K_L|
K_S\rangle$. These analyses started in 1970 and have reached the impressive sensitivity of
$|m(\Kqz)-m(\Kz)|< 4\cdot 10^{-19}$ GeV at 95\% C.L.~in 2012, as presented by
G.~D'Ambrosio at the workshop. An open question remains by how much invisible decays 
of neutral K mesons can influence the result. How well is unitarity tested experimentally?
How much does the product of lifetime and the sum of all measured partial decay rates deviate from unity for $K_S$ and $K_L$ decays? 
And how much would the errors on Re$\epsilon$ and Im$\delta$ increase if the invisible modes would have maximal CP violation?
As long as this is not answered quantitatively by experiments,  ``direct'' tests of
CPT symmetry remain important.\\[5mm]
\noindent
${}^1$ Institut f\"ur Kern- und Teilchenphysik, Technische Universit\"at Dresden, Germany \\ 
${}^2$ Institut f\"ur Kernphysik, Johannes-Gutenberg-Universit\"at Mainz, Germany\\
${}^3$ Max-Planck-Institut f\"ur Physik, M\"unchen, Germany\\
${}^4$ Queen Mary, University of London, United Kingdom\\
${}^5$ Dipartimento di Fisica, Sapienza Universita di Roma, and INFN, Roma, Italy\\
${}^6$ IFIC, Universitat de Valencia-CSIC, Valencia, Spain\\
${}^7$ INFN, Napoli, Italy\\
${}^8$ IPP, ETHZ, Z\"urich, Switzerland\\
${}^9$ Institut f\"ur Experimentelle Kernphysik, Karlsruher Institut f\"ur Technologie, Karlsruhe, Germany\\
${}^{10}$ Justus-Liebig-Universit\"at, Giessen, Germany\\
${}^{11}$ CERN, Geneva, Switzerland\\
\newpage
\section{T Violation in Decays of Neutral B Mesons}
\noindent 
{\bf Klaus R.~Schubert}\\

{\small \noindent \begin{quote} Abstract: The CP-violating observable Im $\lambda$ with 
$\lambda = q\Abar(\Bqz\to J/\psi \Kqz)/pA(\Bz\to J/\psi \Kz)$ can be written as 
$|\Abar/A|\cdot\imlaT$, where $\tilde\lambda = q\Abar|A|/(pA|\Abar|)$. In this product,
$|\Abar/A|$ is CPT violating and $\imlaT$ is T violating. Therefore, observation of 
$\imla\ne 0$ in the $\sin\Delta m t$ 
term of the time-dependent 
rate of $\Bz\to J/\psi K_S$ decays is a proof of T-symmetry violation.
CPT violation in these decays would lead to a non-vanishing $\cos\Delta m t$ term
in the rate. The first measurements of $\imla\ne 0$ in 2001 already prove T
violation. The BABAR 2012 analysis demonstrates that $\imla\ne 0$ leads to a
``motion-reversal'' difference in the rates for the transitions $\Bz,\Bqz\to
B_+,B_-$ and $B_+,B_-\to \Bz,\Bqz$.\end{quote}} 
\vspace{4mm}
\noindent
The notion ``Time Reversal Violation'' has two different meanings, either breaking the
symmetry of the transformation $t\to -t$ in the fundamental laws of an interaction,
or unequal motions (in classical mechanics) or evolutions (in quantum mechanics) for
two systems when exchanging initial and final state and reversing velocities and spins
\cite{Sachs,Schubert-MITP}. I will use the term ``T-symmetry violation'' for the first and
``motion-reversal violation'' for the second case. In particle physics we use
motion-reversal experiments, like the comparison of the two time-dependent rates for
the transitions $\Kz\to\Kqz$ and $\Kqz\to\Kz$, as a method to test T-symmetry of the weak interaction \cite{Ruf-MITP}. Since the recent BABAR experiment 
\cite{2012-BABAR,Bernabeu-MITP} on motion reversal in the transitions $\Bz\to B_-$
and $B_-\to˜\Bz$ has been a central discussion point in this workshop, this note summarizes
its relevance for demonstrating T-symmetry violation.\\

$\Bz\Bqz$ transitions are well known to be sensitive to the symmetries CP, T and CPT.
In the Weisskopf-Wigner approximation, the evolution of the two-dimensional
$B^0$ state $\Psi = \psi_1 B^0 + \psi_2 {\overline B}{}^0$ is given by
$$ i {\dot \psi}_i =\Lambda_{ij}\psi_j,$$
where $\Lambda_{ij}=m_{ij}-i\Gamma_{ij}/2$ with two time-independent 
hermitean 2x2 matrices. The general solution is
$$\psi_i(t)=U_{ij}(t)\psi_j(0)~,$$
where the time-dependent matrix $U_{ij}(t)$ follows 
unambiguously from $\Lambda_{ij}$\cite{BrancoLavouraSilva,FidecaroGerber}.
The transition rates  $|\langle\psi_f|U|\psi_i\rangle|^2$ are determined by
six real observables $\Delta m=m_H - m_L$, $\Gamma =
(\Gamma_H + \Gamma_L)/2$, $\Delta \Gamma =\Gamma_H - \Gamma_L$, 
$${\rm Re}\delta + i~{\rm Im}\delta = \frac{(m_{22}-m_{11})/2 - i (\Gamma_{22}
-\Gamma_{11})/4}{\Delta m - i \Delta\Gamma/2}~{\rm and}~\left|\frac{q}{p}\right|=
\left|\sqrt{\frac{\Lambda_{21}}{\Lambda_{12}}}\right|~.$$
A non-zero phase between $m_{12}$ and $\Gamma_{12}$ leads to $|q/p|\ne 0$, violating 
CP and T; $\Lambda_{11}\ne \Lambda_{22}$ leads to $\delta\ne 0$, violating CP and CPT.
All six observables have been measured \cite{PDG2012}, and $\Delta\Gamma$, 
Re$\delta$, Im$\delta$
and $|q/p|-1$ are compatiblele with zero. Within experimental errors, the Hamiltonian for
$\Bz\Bqz$ transitions is invariant under all transformations T, CPT and CP.\\

Decays of neutral B mesons into $J\psi K_S$ and $J\psi K_L$ are determined by only three
more real observables, if they are dominated by a single weak amplitude,
$$|A|=|\langle J\psi K^0|D|\Bz\rangle|~,~~|\Abar|=|\langle J\psi \Kqz|D|\Bqz\rangle|~,$$
and Im$\lambda$, where $\lambda$ is defined in a phase-convention independent way by
\Begeq \lambda = \frac{q\Abar}{pA}~.\label{Eqlambda}\Endeq
The single-amplitude condition (no ``direct T violation'') is assumed in this summary
and can be tested in the data, as presented later.  
The operator $D$ is the Hamiltonian for the decay, and $|A|^2/\Gamma$ is the fraction of
$\Bz\to J/\psi\Kz$ decays. The relevance of $|\Abar/A|$ and Im$\lambda$ is easily 
seen by introducing the factorization
\Begeq \lambda = \left|\frac{\Abar}{A}\right|\cdot{\tilde\lambda}~{\rm with}~
{\tilde\lambda}=\frac{q\Abar|A|}{pA|\Abar|}~.\Endeq
The observables $|\Abar/A|$ and $\tilde\lambda$ describe the CPT and T
symmetry properties of the decays. If the matrix element $\langle J\psi K^0|D|\Bz\rangle$
is given by a single weak amplitude, then CPT invariance of $D$ requires $|\Abar/A|=1$
\cite{LeeOehmeYang,Gerber-MITP}, T invariance requires $\imlaT =0$
\cite{LewisEnz,Gerber-MITP}, and CP invariance requires $|\Abar/A|=1$ and $\imlaT =0$.
Decays of neutral B mesons into flavour-specific states, e.~g.~into $\ell^\pm\nu X$,
are given by only two more observables,
$$|A_\ell|=|\langle \ell^+\nu X|D|\Bz\rangle|~{\rm and}~
|\Abar_\ell|=|\langle \ell^-\nu X|D|\Bqz\rangle|~.$$
The equivalent to $\imlaT$ is not present if the ``$\Delta Q=\Delta b$'' rule
is strictly valid, i.~e.~
$$\langle \ell^-\nu X|D|\Bz\rangle=\langle \ell^+\nu X|D|\Bqz\rangle =0~.$$

The following text consists of two parts. In the first part I present the consequences
of CPT and T symmetry for the transitions $\Bz,\Bqz\to J/\psi K_S, K_L$ and 
$B_-,B_+\to \Bz,\Bqz$ as derived by H.-J.~Gerber \cite{Gerber-MITP} in collaboration
with M.~Fidecaro and T.~Ruf. In the second part I discuss the consequences for the
2001 analyses of BABAR \cite{2001-BABAR} and Belle \cite{2001-Belle} 
and for the 2012 analysis of BABAR \cite{2012-BABAR}. I will use some additional
simplifying assumptions which have no influence on the main conclusions,
$|q/p|=1$, $\delta=0$, $\Delta\Gamma=0$, 
\Begeq K_S=(\Kz+\Kqz)/\sqrt{2}, ~K_L=(\Kz-\Kqz)/\sqrt{2}~,\label{Eq5a}\Endeq
and validity of the ``$\Delta Q=\Delta b$'' and ``$\Delta S = \Delta b$'' rules in flavour-specific and $J/\psi K$ decays, respectively. This leads to
$$A_S=A(\Bz\to J\psi K_S)=A/\sqrt{2}~,~~\Abar_S=A(\Bqz\to J\psi K_S)=\Abar/\sqrt{2}~,$$
$$A_L=A(\Bz\to J\psi K_L)=A/\sqrt{2}~,~~\Abar_L=A(\Bqz\to J\psi K_L)=-\Abar/\sqrt{2}~.$$
The evolution operator $U(t)$ is given by
\begin{equation} U_{ij}(t)=e^{-\Gamma t/2} \left(\begin{array}{cc}
  \cos (\Delta mt/2) & i\sin (\Delta mt/2)\cdot p/q\\i\sin (\Delta mt/2)
  \cdot q/p & \cos (\Delta mt/2)\end{array}\right)~,\label{Eq1}\end{equation}
and $\Delta S = \Delta b$ means that the decay matrix is diagonal,
\Begeq D=\left(\begin{array}{cc}A(B^0\to J/\psi K^0) & A(\Bqz\to J/\psi K^0)\\
A(B^0\to J/\psi \Kqz) & A(\Bqz\to J/\psi {\overline K}{}^0)\end{array}\right)
  =\left(\begin{array}{cc}A & 0\\ 0 & {\overline A}\end{array}\right)~. 
  \label{Eq4}\Endeq

The decay rate $R=R(B^0\to J/\psi K_S|t)$ of a $\Bz$ at time $t=0$ into the final 
state $J/\psi K_S$ at time $t$ is
$$R= \frac{1}{2} e^{-\Gamma t}\left|\left(\begin{array}{cc}A 
   & \Abar\end{array}\right)    
   \left(\begin{array}{cc}\cos (\Delta mt/2) & i \sin (\Delta mt/2)\cdot p/q\\
   i \sin (\Delta mt/2)\cdot q/p & \cos (\Delta mt/2)
   \end{array}\right)\left(\begin{array}{c} 1 \\ 0\end{array}\right)\right|^2$$
\Begeq = |A|^2\cdot
   \frac{e^{-\Gamma t}}{2}  
   \left[1+\kappa-\kappa\cdot\cos(\Delta m t)-{\rm Im}\lambda\cdot
   \sin(\Delta m t) \right]~,\label{Eq5}\Endeq
with the CPT-violating parameter 
\begin{equation}\kappa =\frac{|\lambda|^2-1}{|\lambda|^2+1}\label{Eq6}\end{equation}
and Im$\lambda$ from eq.~\ref{Eqlambda}. I also
assume $\kappa \ll 1$, i.~e.~$|\lambda|^2=1+2\kappa$. If CPT is conserved, $\kappa=0$
and the cosine term has to vanish. If T is conserved, Im$\lambda=0$ and the sinus
term has to vanish. If CP is conserved, the rate has to be a pure exponential.
With good statisics and well-understood
systematics, tests of CP, T and CPT can be made with a single time-dependent
measurement of the rate for $B^0\to J/\psi K^0_S$ decays. Additional rate measurements of 
$B^0\to J/\psi K^0_L$, ${\overline B}{}^0\to J/\psi K^0_S$ and ${\overline B}{}^0
\to J/\psi K^0_L$ improve statistics as well as systematic uncertainties; but the full physics information is already contained in the time
dependence of $B^0\to J/\psi K^0_S$ decays alone.\\

The derivation of the rate in eq.~\ref{Eq5} allows a simple argument for the fact that
Im($q{\overline A}/pA) \ne 0$ proves T violation of the Hamiltonian \cite{Gerber}. 
Im($q{\overline A}/pA) \ne 0$ violates CP symmetry which implies CPT and/or T violation.
CPT symmetry requires only $|{\overline A}/A|=1$ and nothing for the phase of 
${\overline A}/A$. Since Im$(q{\overline A}/pA) \ne 0$ does not contradict CPT symmetry, 
it must violate T. The same conclusion that $\imla\ne 0$ is CP- and T-violating is 
found by P.~Villanueva-Perez \cite{Pablo}.\\

The rates for $B^0\to J\psi K^0_L$, ${\overline B}{}^0\to J\psi K^0_S$ and ${\overline B}{}^0\to J\psi K^0_L$ follow from replacing $(A,\Abar)$
and $(1,0)$ in eq.~\ref{Eq5} by $(A,-\Abar)$ and/or $(0,1)$. 
All four rates are
given by the same expression 
\Begeq R = |A|^2\cdot\frac{e^{-\Gamma t}}{2}\left[1+\kappa +C\cdot\cos(\Delta m t)+S\cdot
\sin(\Delta m t) \right]~,\label{Eq5b}\Endeq
with different values for the coefficients $C$ and $S$ as given in 
Table \ref{tab1}.\\

\begin{table}[h]\begin{center}
\caption{Coefficients $C$ and $S$ for the rates in eq.~\ref{Eq5b}.}\label{tab1} \vspace{0.2cm}
\begin{tabular}{|l|c|c|}\hline
 & $C$ & $S$\\ \hline
$B^0\to J\psi K^0_S$  & $-\kappa$ & $-\imla$ \\
$B^0\to J\psi K^0_L$  & $-\kappa$ & $+\imla$\\
$\Bqz\to J\psi K^0_S$ & $+\kappa$ & $+\imla$ \\
$\Bqz\to J\psi K^0_L$ & $+\kappa$ & $-\imla$\\ 
\hline\end{tabular}\end{center}\end{table}

The T violation in all four rates is given by the same parameter $\imla$,
the CPT violation in all four by $\kappa$. From
entangled $\Bz\Bqz$ pairs with tagging by
flavour-specific decays, the $B^0$ rates obtain an extra factor
$|\Abar_\ell|^2$, the $\Bqz$ rates $|A_\ell|^2$. The four rates are only equal, up to the signs in the table, if $|\Abar_\ell|=|A_\ell|$, i.~e.~CPT symmetry in flavour-specific
decays.\\

With the Kaon sign-conventions 
in eq.~\ref{Eq5a}, the states $B_+$ and $B_-$ \cite{Bernabeu,Alvarez,BMV}, 
\Begeq B_+ = N(\Abar B^0 - A\Bqz)~,~~ B_- = N (\Abar B^0 + A\Bqz)~,~~
N^{-2}= |A|^2 + |\Abar|^2=2|A|^2(1+\kappa)\label{Eq8}\Endeq
have the properties that $B_+$
decays only into $J/\psi K^0_L$, not into $J/\psi K^0_S$, and $B_-$ only into $J/\psi K^0_S$, 
not into $J/\psi K^0_L$.
Since the amplitudes in eq.~\ref{Eq4} are defined as $A=\langle J/\psi \Kz|D|\Bz\rangle$
and $\Abar=\langle J/\psi \Kqz|D|\Bqz\rangle$, the states in eq.~\ref{Eq8} are
ingoing states $|B_+\rangle$ and $|B_-\rangle$ and not the outgoing states in the
transitions $\Bz,\Bqz\to B_+, B_-$. With direct CPT violation, the two states are not
orthogonal, $\langle B_+|B_-\rangle =\kappa$. The orthogonal states to $B_+$ and
$B_-$ are
\Begeq B_{+\perp}=N(A^*\Bz+\Abar{}^*\Bqz)~{\rm and}~B_{-\perp}=N(A^*\Bz-\Abar{}^*\Bqz)~,
\Endeq
respectively, and the normalization factor $N$ is the same as in eq.~\ref{Eq8}. 
Independent of CPT invariance, an $\Upsilon(4S)$ meson decays into the entangled 
two-body state
\Begeq B^0\Bqz - \Bqz  B^0 = B_+ B_{+\perp} - B_{+\perp} B_+ =
                             B_- B_{-\perp} - B_{-\perp} B_-~,\Endeq
where the first $B$ in this notation moves in direction $\vec p$ and the second one
in direction $-\vec p$.
With CPT invariance, $B_{+\perp}=B_-$ and $B_{-\perp}=B_+$. With $\kappa\ne 0$, 
$B_{+\perp}$ decays into both $J\psi K_S$ and $J\psi K_L$. But if the first
$B$ decay of the entangled pair is $J\psi K_S$, then the remaining $B$ is in the
state $B_+$. The state $B_{+\perp}$ is not used in the following motion-reversal
discussion, its only purpose is the preparation of the state $B_+$. The analoguous
argument holds for $B_{-\perp}$. The decays $\Upsilon(4S)\to\Bz\Bqz\to
(J/\psi K_S)(J/\psi K_S)$  and $(J/\psi K_L)(J/\psi K_L)$
are forbidden in accordance with Bose statistics.
As a side remark, not relevant for T violation, the two states $B_+$ and $B_-$ are well defined physical states like the mixing eigenstates $B_H$ and 
$B_L$, free of phase conventions, but all four are not CP eigenstates.
The decay rates of $B_+$ and $B_-$ are
\Begeq \Gamma_{\psi K}=\Gamma(B_-\to J\psi K_S)=\Gamma(B_+\to J\psi K_L)=|A|^2(1+\kappa)~,\label{Eq10}\Endeq
where the factor $1+\kappa$ originates from the normalization $N$ in eq.~\ref{Eq8}. Dividing the rates $R(t)$ in eq.~\ref{Eq5} by this decay rate leads to
\Begeq R_1(t)=\frac{e^{-\Gamma t}}{2}\left[1 +C_1\cdot\cos(\Delta m t)+S_1\cdot
\sin(\Delta m t) \right]~,\label{Eq11}\Endeq
with the coefficients $C_1$ and $S_1$ as given in Table \ref{tab2}, using
$\imlaT=\imla/(1+\kappa) =\imla(1-\kappa)$.
Note that these rates $R_1(t)=R(t)/\Gamma_{\psi K}$ are not the rates for the transitions
$\Bz,\Bqz\to B_+, B_-$ because of the difference between in- and outgoing states.\\

Calculation of the time-dependent rates $R_2(t)$ for the transitions 
$\Bz,\Bqz\to B_+, B_-$ requires the outgoing states
\Begeq \langle B_+| = N(\Abar^* \langle B^0| - A^* \langle\Bqz|)~,~~
       \langle B_-| = N(\Abar^* \langle B^0| + A^* \langle\Bqz|)~, 
\label{Eq8f}\Endeq
leading to
\Begeq R_2(t)=\frac{e^{-\Gamma t}}{2}\left[1 +C_2\cdot\cos(\Delta m t)+S_2\cdot
\sin(\Delta m t) \right]~,\label{Eq12}\Endeq
with $C_2$ and $S_2$ in Table~\ref{tab2}. The motion-reversed transitions
$B_+,B_-\to\Bz,\Bqz$ require the ingoing states of $B_+$ and $B_-$. Their rates
$R_3(t)$ are found to be
\Begeq R_3(t)=\frac{e^{-\Gamma t}}{2}\left[1 +C_3\cdot\cos(\Delta m t)+S_3\cdot
\sin(\Delta m t) \right]~,\label{Eq12mr}\Endeq
with $C_3$ and $S_3$ in Table~\ref{tab2}.\\ 

\begin{table}[h]\begin{center}
\caption{Coefficients $C_i$ and $S_i$, $i=1$ for the rates 
$R(B\to J/\psi K)/\Gamma_{\psi K}$ in eq.~\ref{Eq11},
$i=2$ for the transitions in eq.~\ref{Eq12}, and $i=3$ for the transitions in 
eq.~\ref{Eq12mr}.}\label{tab2} \vspace{0.2cm}
\begin{tabular}{|l|c|c|c|c|c|c|c|c|}\hline
 & $C_1$ & $S_1$ & & $C_2$ & $S_2$ & & $C_3$ & $S_3$\\ \hline
 &       &       & &       &       & &       &      \\
$B^0\to J/\psi K_S$  & $-\kappa$ & $-\imlaT$ & $\Bz\to B_-$ & $+\kappa$ & $-\imlaT$
              & $B_-\to\Bz$ & $+\kappa$ & $+\imlaT$ \\
$B^0\to J/\psi K_L$  & $-\kappa$ & $+\imlaT$ & $\Bz\to B_+$ & $+\kappa$ & $+\imlaT$
              & $B_+\to\Bz$ & $+\kappa$ & $-\imlaT$ \\
$\Bqz\to J/\psi K_S$ & $+\kappa$ & $+\imlaT$ & $\Bqz\to B_-$ & $-\kappa$ & $+\imlaT$
              & $B_-\to\Bqz$ & $-\kappa$ & $-\imlaT$ \\
$\Bqz\to J/\psi K_L$ & $+\kappa$ & $-\imlaT$ & $\Bqz\to B_+$ & $-\kappa$ & $-\imlaT$
              & $B_+\to\Bqz$ & $-\kappa$ & $+\imlaT$ \\ 
\hline\end{tabular}\end{center}\end{table}

The time-dependent motion-reversal asymmetries like between $\Bz\to B_-$ and
$B_-\to\Bz$,
\Begeq {\cal A}_{MR} = \frac{R_3(t)-R_2(t)}{R_3(t)+R_2(t)}=
   \frac{C_3-C_2}{2}\cdot\cos(\Delta m t)
   +\frac{S_3-S_2}{2}\cdot\sin(\Delta m t)~,\label{Eq13}\Endeq
are sensitive to only T-symmetry violation; $(C_3-C_2)/2=0$
and $(S_3-S_2)/2=S_3=\pm \imlaT$.
The time-dependent quasi-motion-reversal asymmetries like between $\Bz\to J/\psi K_S$ and
$B_-\to\Bz$, 
\Begeq {\cal A}_{QMR} = \frac{R_3(t)-R_1(t)}{R_3(t)+R_1(t)}=
   \frac{C_3-C_1}{2}\cdot\cos(\Delta m t)
   +\frac{S_3-S_1}{2}\cdot\sin(\Delta m t)~,\label{Eq13-2}\Endeq
are sensitive to both CPT- and T-symmetry violation; $(C_3-C_1)/2=C_3=\pm\kappa$
and $(S_3-S_1)/2=S_3=\pm \imlaT$.\\

The CP-violation analyses \cite{2001-BABAR,2001-Belle} have combined the
transitions $\Bz,\Bqz\to J\psi K_S, K_L$ and $B_-, B_+\to \Bz,\Bqz$
by using events with both signs of $\Delta t = t({\rm decay~to}~J/\psi K_{S,L})
- t({\rm decay~to}~\ell^\pm X)$. Recalling that
$\Delta t=+t$ and $\Delta t=-t$ for the events with $i=1$ and $i=3$ in Table 
\ref{tab2} respectively, and inspecting the signs of $C_1$ and $C_3$, shows that the determination 
of $|\lambda|$ in these analyses is also a determination of
$\kappa$. Accidental cancellations between direct T violation (two decay amplitudes
with different phases giving leading to $|\lambda|\ne 1$) and direct CPT violation
($\kappa\ne 0$) are very unlikely. Therefore, the analyses determine $\imlaT$ and 
$\kappa$ and 
show T-symmetry violation with a significance of about $4\sigma$ independent 
of any assumption on CPT \cite{Gerber-MITP}.\\ 

The recent BABAR analysis \cite{2012-BABAR} has for the first time separated the
CP-violation data into events with $\Delta t >0$ and $\Delta t<0$ and has determined 
eight uncorrelated coefficients $C_i$ and $S_i$ for the $\cos\Delta m t$
and  $\sin\Delta m t$ terms. All $C_i$ values are compatible with CPT symmetry, and all
$S_i$ values  prove T-symmetry violation. Since earlier analyses can lead to the same conclusion, the main merit of the analysis in ref.~\cite{2012-BABAR} is its 
demonstration of T-symmetry violation
by motion-reversal violation. The data can only use 
quasi-motion-reversals, comparing e.~g.~$\Bz\to J/\psi K_S$ and $B_-\to B^0$, and not
motion-reversals with $\Bz\to B_-$ and $B_-\to B^0$. However, the null results for the eight determinations of $\kappa$ justify this approximation.

\newpage % ===================================================================================

\section{T and CPT studies in $B^0 \bar{B}^0$ transition with Belle}
\setcounter{equation}{0}
\noindent 
{\bf Luigi Li~Gioi}\\

The Belle collaboration, having performed experiments at the KEKB B-factory since 1999, made the essential observations of the CP violation, proving the differences in the decays of B mesons compared to their anti-particles $\bar{B}^0$.\\

The KEKB B-factory is currently under upgrade to a new generation of super flavor factory (SuperKEKB) which aims to deliver more than 50 $ab^{-1}$ by the end of 2022.
Together with the machine, the Belle detector upgrade is ongoing (Belle II)~\cite{ref:BELLE2}. The main redesign goals are to cope with the much 
higher physics rates and the much larger background, as well as improving the overall physics performance.
For T and CPT violation studies the new tracking system plays a key role.
The tracking system in the former Belle detector consisted of 4 layer of Si strip vertex detector (SVD), followed by a Central Drift Chamber (CDC).
For the new tracking system a two-layer pixel detector (PXD) for the innermost Si layers will be implemented. The SVD will be replaced entirely as well as the CDC: due
to the harsh backgrounds the inner radius of the CDC has to be moved out and the two outer layers of the new SVD will cover the gap.
The momentum resolution of charged particles will be improved by extending the CDC to a larger radius.\\

The impact parameters: $d_0$ and $z_0$, defined as the projections of the distance from the point of closest approach to the origin, are a good measure of the
overall performance of the tracking system and as such are used to find the optimal tracker configuration. An improvement of roughly a factor two is expected on
the impact parameter resolution. The introduction of a new vertex fitter together with the improvement of the alignment procedure will 
thus sensibly improve the systematic error of any time dependent measurement. 

\subsection{CPT Violation}

In the presence of CPT violation, the flavor and mass eigenstates of the neutral B mesons are related by $|B_L> = p \sqrt{(1-z)} |B^0> + q \sqrt{(1+z)} |\bar{B}^0>$
and $|B_H> = p \sqrt{(1+z)} |B^0> - q \sqrt{(1-z)} |\bar{B}^0>$ where $|B_L>$ and $|B_H>$ are the light and heavy mass eigenstates.
Here $z$ is a complex parameter accounting for CPT violation; CPT is violated if $z \neq 0$.
Then, the time dependent decay rate of the two B mesons generated from the $\Upsilon(4S) \to B^0 \bar{B}^0$ is given by~\cite{ref:DTCPT}:
\begin{small}
\begin{multline}
P(\Delta t, f_{rec},f_{tag})=  \frac{\Gamma_d}{2} e^{-\Gamma_d |\Delta t|} [ \frac{|\eta_+|^2 + |\eta_-|^2}{2} 
cosh\left( \frac{\Delta\Gamma_d}{2} \Delta t \right) -  Re(\eta_+^* \eta_-) sinh\left( \frac{\Delta\Gamma_d}{2} \Delta t \right)  \\
  +  \frac{|\eta_+|^2 - |\eta_-|^2}{2} cos(\Delta m_d \Delta t) + Im(\eta_+^* \eta_-) sin(\Delta m_d \Delta t) ]  
\end{multline}
\end{small}
where $\eta_+= A_1 \bar{A}_2 - \bar{A}_1 A_2$, $\eta_-=\sqrt{(1-z^2)}(p/q A_1 A_2 - q/p \bar{A}_1 \bar{A}_2) + z(A_1 \bar{A}_2 + \bar{A}_1 A_2)$,
$A_1$ and $A_2$ are the decay amplitudes of the reconstruction and tag side B mesons to $f_1$ and $f_2$ final states.\\

The Belle collaboration measured the CPT-violating parameter $z$ and the normalized total decay-width difference $\Delta\Gamma_d/\Gamma_d$~\cite{ref:CPT} in
$B^0 \to J/\psi K^0$ ($K^0 = K_S,K_L$), $B^0 \to D^{(*)-} h^+$ ($h^+=\pi^+$ for $D^-$ and $\pi^+,\rho^+$ for $D^{*-}$), and $B^0 \to D^{*-} l^+ \nu_l$
($l^+=e^+,\mu^+$) decays. Most of the sensitivity to $Re(z)$ and $\Delta\Gamma_d/\Gamma_d$ is obtained from neutral B-meson decays to $f_{CP}$, 
while $Im(z)$ is constrained primarily from other neutral B-meson decay modes.
The results are based on a data sample of $535 \times 10^6 B\bar{B}$ pairs collected at the $\Upsilon(4S)$ resonance: 
$Re(z)=[+1.9 \pm 3.7(stat) \pm 3.3(syst)]\times 10^{-2}, Im(z)=[-5.7 \pm 3.3(stat) \pm 3.3(syst)]\times 10^{-3}$ and 
$\Delta\Gamma_d/\Gamma_d= [-1.7 \pm 1.8(stat) \pm 1.1(syst)] \times 10^{-2}$, all of which are consistent with zero. This is the
most precise single measurement of these parameters in the neutral B-meson system to date.
The dominant contributions to the systematic uncertainties are from vertex reconstruction and the tag-side interference (TSI)~\cite{ref:TSI};
the next largest contributions are from fit bias. Using the new tracking system of the Belle II detector and refining the analysis will then permit to
sensibly reduce the amount of the systematic uncertainty and to fully exploit the larger data sample that will be collected by the Belle II experiment. 
Assuming a final integrated luminosity of 50 $ab^{-1}$, an improvement of $\sqrt{50/0.5}=10$ is expected.

\subsection{T Violation}

The CPLEAR collaboration reported on the observation of time-reversal symmetry violation through a comparison of the probabilities of $K^0$
transforming into $\bar{K}^0$ and $\bar{K}^0$ into $K^0$ as a function of the neutral-kaon eigentime $t$. An average decay-rate asymmetry 
$<A_T> = [6.6 \pm 1.3(stat) \pm 1.0(syst)] \times 10^{-3}$ was measured over the interval $1\tau_S < \tau < 20\tau_S$~\cite{ref:TCPLEAR}.\\
 
In the case of the $B^0 \bar{B}^0$ mesons the asymmetry is then expected to be close to zero; 
a value significantly larger than $10^{-3}$ would be an indication of new physics~\cite{ref:ASLNP}.\\

The Belle collaboration performed this measurement using the semileptonic decays of the neutral $B$ meson~\cite{ref:ASLBELLE}: 
a possible difference between the transitions $B^0 \to \bar{B}^0$
and $\bar{B}^0 \to B^0$ can manifest itself as a charge asymmetry in same-sign dilepton events in $\Upsilon(4S)$ decays when prompt leptons from
semileptonic decays of neutral B mesons are selected. Using a data sample of 78 $fb^{-1}$ recorded at the $\Upsilon(4S)$ resonance and 9 $fb^{-1}$ recorded at an energy 
60 MeV below the resonance, Belle measures $A_{sl} = [1.1 \pm 7.9(stat) \pm 8.5(syst) ] \times 10^{-3}$.
The dominant contributions to the systematic uncertainties are from Track finding efficiency and Continuum subtraction; they can be reduced with a better knowledge of the
tracking system and a refinement of the analysis.\\

Recently the BaBar collaboration reported a measurement of T-violating parameters in the time evolution of neutral B mesons~\cite{ref:TBABAR} using the decays 
of entangled neutral B mesons into definite flavor states ($B^0$ or $\bar{B}^0$), and $J/\psi K_L$ or $c\bar{c}K_S$ final states with the comparisons between the 
probabilities of four pairs of T-conjugated transitions as a function of the time difference between the two B decays~\cite{ref:BERNABEU}.
The results obtained by the BaBar collaboration are in agreement with CPT conservation.\\

At this point the Belle collaboration would not expect that repeating the BaBar analysis would improve the results significantly.
However there is an interest in these results as a test of CPT.\\

The experimental method consists in dividing the very same $\Delta t$ distribution used in the standard CP analysis in two parts: 
$\Delta t > 0$ and  $\Delta t < 0$, and fitting them using the same function used for the CP violation analysis:
\begin{equation}
g_{\alpha,\beta}^{\pm} (\Delta\tau) = e^{-\Gamma_d \Delta\tau} \{1 + S_{\alpha,\beta}^{\pm} sin(\Delta m_d \Delta\tau) + C_{\alpha,\beta}^{\pm} cos(\Delta m_d \Delta\tau) \}
\end{equation}
where $\Delta\Gamma_d = 0$ is assumed, the indices $\alpha=l^+l^-$ (flavor state), $\beta=K_S,K_L$ (CP states) and the symbol $+$ or $-$ indicates whether the decay to the 
flavor final state $\alpha$ occurs before or after the decay to the CP final state $\beta$. $\Gamma_d$ is the average decay width and $\Delta m_d$ is the
mass difference between the neutral B mass eigenstates. Since, using the data sample collected by the Belle experiment in the standard CP violation analysis~\cite{ref:CPBELLE},
no tension is observed in the fit of the $\Delta t$ distributions between the left and right sides of the distributions and between $J/\psi K_L$ and $c\bar{c}K_S$,  
this analysis could not yield results compatible with CPT violation.\\ 

The measurement of the T-violation parameters could become important for the Belle II collaboration, when a very high experimental precision will be achieved. 
At that time also a number of measurements should be accessible to test T symmetry invariance in $b \to u$,$d$ and $s$ transitions as well as in the charm sector 
to test $c \to d$ and $c \to s$ transitions~\cite{ref:TOTHER}. It is also considered an advantage that any CP and T measurement should yield about the same result
if CPT is conserved. This would allow a cross check of any CP measurement.

\newpage % ===================================================================================
%\vspace{8mm}
\section{Future Measurements of T violation in B and D decays}
\setcounter{table}{0}
\noindent 
{\bf Adrian J.~Bevan}\\

{\small \noindent \begin{quote} Abstract: The potential for future measurements of \T violation in \B and \D decays is summarised here.
This discussion considers possible quark flavour changing transitions from $b$ and $c$ quarks to all 
kinematically accessible final states via tree and loop topologies.\end{quote}} 
\vspace{2mm}
\noindent
There is scope to extend the tests of the \T symmetry using quantum entangled pairs of neutral mesons
outlined in Refs~\cite{Banuls:1999aj,Banuls:2000ki,Bernabeu:2012ab}, and performed recently by
\babar~\cite{Lees:2012kn}, to other \CP filter final states of \B decays as well as for some charm 
decays.  These measurements can be used to 
test the nature of the \T symmetry under quark transitions at tree and loop level in 
$b\to c$, $s$, $u$, and $d$ transitions, as well as $c\to d$ and $s$ transitions. 
The loop transition $c\to u$ is experimentally inaccessible given that the corresponding
loop is Cabibbo suppressed (i.e. by a factor of $\Vub \propto \lambda^3$) and more copious
loop transitions will dominate any attempt to extract that term. There
are two categories of orthonormal \CP basis filters that can be used for such measurements:
(i) the approximately orthonormal pair of $T$-conjugate decays $X+\KS$ and $X+\KL$,
and (ii) \T self-conjugate decays of a pseudoscalar to two spin-one particles where
a transversity analysis of the final state allows one to experimentally distinguish
between \CP even and \CP odd parts of the decay.  This summary is based on
Ref.~\cite{Bevan:2013rpr} and naive numerical estimates prepared 
for this workshop. These estimates are obtained using existing experimental results 
on time-dependent \CP asymmetry measurements from the \BFs.  Only uncertainties are
quoted as it would be incorrect to extrapolate central values for the \CP asymmetry parameters
in terms of the \T asymmetry parameters ${\rm\Delta} S_T^\pm$ without re-analysing the data. 

\subsection{$B$ decay measurements at the \FourS}

The \babar and \belle experiments already have data that can be used to perform a number of
\T violation tests beyond the $b\to c$ transitions described in Ref.~\cite{Lees:2012kn}.  In
the near future \belle II is expected to accumulate 50\invab of data which will enable one
to perform precision measurements of time-dependent $T$-violating asymmetries that can be 
related to the CKM matrix description of quark flavor changing transitions.  \T violation
has been established in entangled $B$ meson decays to combinations of flavor tagged
final states denoted by $\ell^\pm X$ and \CP filter final states $B_+$ and $B_-$, taken
to be $c\overline{c} K_{S,L}$.  The presence of two sets of orthonormal filters enables
one to test \T, \CP and \CPT via the different pairings of events as a function of 
proper time difference $\rm \Delta \tau$ as outlined by Bernabeu et al.  
%A \Bz decays into
%a final state with a \Kz, and a \Bzb decays into a final state with a \Kzb, where the strong
%states can be related to the weak states \KS and \KL via a linear combination.  
In analogy
with the $b\to c$ transitions proposed by Bernabeu et al., it is possible to study \T-violation
in the interference between mixing and decay amplitudes in $b\to s$ transitions involving 
$\eta^\prime K_{S, L}$, $\phi K_{S, L}$ and $\omega K_{S, L}$
\CP filter pairs, however the \BFs only have sufficient data for the first two measurements as
$B\to \omega \KL$ has not yet been observed.  Nonetheless one can estimate the anticipated
precision attainable on a test of \T using the $\omega K_{S,L}$ \CP filters.\\

The set of $B_+$ and $B_-$ involving $\KS$ and $\KL$ is approximately orthonormal (this is a good approximation given
current levels of experimental precision).  As mentioned above it is also possible to define an exactly orthonormal
basis pair for the $B_+$ and $B_-$ filters as illustrated in the following. 
The $b\to c$ transition $\B\to J/\psi K^{*0}$ 
is composed of three $P$ wave parts, one is 
a longitudinal component that is \CP even and there are two transverse components: the 
perpendicular (\CP odd) and parallel (\CP even) parts.  As a result it is possible to 
define an exactly orthonormal basis of $B_+$ and $B_-$ decays into the $J/\psi K^{*0}$
final state (the same is true for other decays of a neutral pseudo-scalar 
meson to two spin-one particles).  In analogy 
with this discussion, one can also test the \T symmetry using the \CP even and odd
parts of other modes, such as the decay $B\to D^*D^*$ (a $b\to d$ quark transition),
$B\to \phi K^*$ (a $b\to s$ quark transition), and $B\to \rho^0\rho^0$ 
(a $b\to u$ quark transition) as a basis of \CP filters.\\

Estimates of the experimental precisions attainable for these modes, where possible to compute, can be found in 
Table~\ref{tbl:bfactoryestimates}.  It is likely that Belle II will be able to observe
\T violation at a significance of at least $5\sigma$ for all of the modes listed, assuming
that the violation occurs at the level expected within the SM.  Given that 
${\rm\Delta} S_T^\pm$ is related to $\lambda_f = (q/p)(\overline{A}/A)$, one can relate
these modes to the angles of the Unitarity triangle, and in this case (for
$b\to c$, $d$, and $s$ transitions) one has
a set of tree and loop dominated measurements of ${\rm\Delta} S_T^\pm = \mp 2\sin 2\beta \simeq \mp 1.36$,
where the cosine coefficient ${\rm\Delta} C_T^\pm$ is expected to be zero in the 
SM.  The decay $B\to \rho^0\rho^0$ can be used to measure the Unitarity triangle
$\alpha$ in the SM.  There is insufficient experimental data to extrapolate the 
precision with which one may be able to constrain ${\rm\Delta} S_T^\pm$ in 
$B\to \rho^0\rho^0$, however such a measurement is feasible at Belle II.

\begin{table}[!ht]
\caption{Estimates of the precisions on the \T asymmetry parameter 
${\rm\Delta} S_T^\pm$ for different experiments, based on existing 
results from the \BFs.}\label{tbl:bfactoryestimates}
\renewcommand{\arraystretch}{1.2}
\begin{center}
\begin{tabular}{l|cc}
\hline \hline
\CP filters        & $\sigma({\rm\Delta} S_T^\pm)$ \babar / \belle & $\sigma({\rm\Delta} S_T^\pm)$ \belle II \\ \hline
$\eta^\prime \Kz$  & 0.56 & 0.08 \\
$\phi K^*$         & 1.14 & 0.13 \\
$\phi \Kz$         & 1.84 & 0.17 \\
$\omega \Kz$       & 1.95 & 0.22 \\
$D^*D^*$           & 2.0  & 0.29 \\
\hline \hline
\end{tabular}
\end{center}
\end{table}

\subsection{$D$ decay measurements at an asymmetric energy $\tau$-charm experiment}

A number of time-dependent \CP asymmetry measurements in charm have been proposed
in order to measure the phase of mixing, and to constrain the $cu$ Unitarity triangle
angle $\beta_c$ (which should be small in the Standard Model)~\cite{Bevan:2011up}.
Neutral $D$ meson pairs created at the charm threshold, $\psi(3770)$, decay
in an entangled state in analogy with neutral \B mesons at the \FourS. As a result
there are a number of $D\to X K_{S,L}$ states and a number of final states
composed of two spin-one particles that can be used to test the \T symmetry in the charm sector.  These can be classified
into those states that measure the phase of charm mixing (via a $c\to s$ transition),
and those that measure a combination of the mixing phase and $\beta_c$ (via
a $c\to d$ transition).  Decays of the first type include $\phi K_{S,L}$ and $\pi^+\pi^- K_{S,L}$
final states, whereas decays of the second type include $K^+K^- K_{S,L}$ and $\rho^0\rho^0$ final 
states.

\newpage %==================================================================
\section{Direct test of T and CPT symmetries in the entangled neutral kaon system at a $\phi$-factory}
\setcounter{equation}{0} \setcounter{table}{0}
\noindent 
{\bf Antonio Di Domenico}\\

\subsection{Introduction}
In this note the possibility to perform direct tests of T and CPT symmetries in the neutral kaon system is discussed. Here {\it direct test} means a test whose outcome 
is independent from the result of any other discrete symmetry test, as discussed in Refs.\cite{ref:Wolfenstein,ref:bernabeuPLB-NPB,ref:BernabeuDiscrete,ref:Bmethod,tviol}.
In order to implement such direct tests
it has been suggested to exploit the 
Einstein-Podolsky-Rosen (EPR) 
entanglement of neutral mesons produced at a $\phi$-factory (or B-factory) \cite{ref:bernabeuPLB-NPB,ref:BernabeuDiscrete,ref:Bmethod,tviol}.
In fact in this case
the initial state of the neutral kaon pair produced in $\phi\rightarrow \kn\knb$ decay can be rewritten in terms of any pair of orthogonal states
$|\kpp \rangle$ and $|\knn \rangle$:
%, e.g. the CP eigenstates:
\begin{eqnarray}
  |i \rangle   =  \frac{1}{\sqrt{2}} \{ |\kn \rangle |\knb \rangle - 
 |\knb \rangle |\kn \rangle
\} 
\label{eq:state1}
   =  \frac{1}{\sqrt{2}} \{ |\kpp \rangle |\knn \rangle - 
 |\knn \rangle |\kpp \rangle
\label{eq:state3}
\}~.
\end{eqnarray}  
Here one can consider the states $|\kpp\rangle$, $|\knn\rangle$ defined as follows:
$|\kpp\rangle$ is the state filtered by the decay into $\pi\pi$ 
($\pi^+\pi^+$ or $\pi^0\pi^0$), a
pure $\CP=+1$ state; analogously $|\knn\rangle$ is the state filtered by the decay into 
$3\pi^0$, a
pure $\CP=-1$ state. 
%With these definitions of $|\kpp\rangle$ and $|\knn\rangle$,
Their orthogonal states correspond to the states which cannot decay into 
$\pi\pi$ or $3\pi^0$, defined, respectively, as
\begin{eqnarray}
|\knnp\rangle &\propto&  \left[| \kln\rangle 
- \eta_{\pi\pi}|\ksn \rangle \right] \nonumber \\
|\kppp\rangle &\propto& \left[| \ksn\rangle 
- \eta_{3\pi^0} |\kln \rangle \right]~, 
\end{eqnarray}
with
$\eta_{\pi\pi}={\langle \pi\pi |T |\kln\rangle}/{\langle \pi\pi |T |\ksn\rangle}$
and 
$\eta_{3\pi^0}={\langle 3\pi^0 |T |\ksn\rangle}/{\langle 3\pi^0 |T |\kln\rangle}$.
With these definitions of states, it can be shown that 
the condition of orthogonality $\langle\knn|\kpp\rangle=0$, 
(i.e. $|\kpp\rangle\equiv|\kppp\rangle$ and
$|\knn\rangle\equiv|\knnp\rangle$)
corresponds to assume negligible direct CP (or CPT) violation contributions (i.e. 
$\epsilon^{\prime}, \epsilon^{\prime}_{000} \ll \epsilon$),
assumption quite well satisfied for neutral kaons 
(see detailed discussion in Appendix A 
of Ref. \cite{tviol}).
The validity of the $\Delta S=\Delta Q$ rule is also assumed, so that the two flavor orthogonal eigenstates $|\kn\rangle$ and $|\knb\rangle$ are identified by the charge 
of the lepton in semileptonic decays, i.e. a $|\kn\rangle$ can decay into $\pi^-\ell^+\nu$ and not into $\pi^+\ell^-\bar{\nu}$, and vice-versa for a $|\knb\rangle$.\\

Thus, exploiting the perfect anticorrelation of the states implied 
by Eq.~(\ref{eq:state3}),
 it is possible to have a 
\textquotedblleft flavor-tag\textquotedblright or a 
 \textquotedblleft CP-tag\textquotedblright,
% ~\cite{ref:bernabeuJHEP},
i.e.~to infer the flavor ($\kn$ or $\knb$) or the CP ($\kpp$ or $\knn$) state
of the still alive kaon by observing a specific flavor decay ($\pi^+\ell^-\nu$
or  $\pi^-\ell^+\bar{\nu}$) or CP decay ($\pi\pi$ or $\pi^0\pi^0\pi^0$) of the other (and first decaying) kaon in the pair.
In this way 
one can experimentally access the transitions listed in Table~\ref{tab:processes},
which
can be divided
into four categories of events, corresponding to independent T, CP and CPT tests.
%as summarized .
\begin{table}[h]
  \begin{center}
    \begin{tabular}{c|c|c|c}		
      \hline
%      \multicolumn{2}{c|}{Reference}  &   \multicolumn{2}{c}{\T-conjugate} \\
      Reference & T-conjug.   &  CP-conjug. & CPT-conjug. \\ \hline
%      \trule 
$\kn \to \kpp$   & $\kpp \to \kn$    & $\knb \to \kpp$ & $\kpp \to \knb$ \\
%      \trule 
$\kn \to \knn$ &  $\knn \to \kn$ &  $\knb \to \knn$ &  $\knn \to \knb$ \\
%     \trule 
$\knb \to \kpp$   & $\kpp \to \knb$    &  $\kn \to \kpp$ & $\kpp \to \kn$ \\
%      \trule 
$\knb \to \knn$  & $\knn \to \knb$   & $\kn \to \knn$  & $\knn \to \kn$   \\ \hline
   \end{tabular}
    \caption{Scheme of possible reference transitions and their associated
 T, CP or CPT conjugated processes accessible at a $\phi$-factory.
%    time-ordered decay products in the experimental $\phi$-factory scheme.
      \label{tab:processes}}
  \end{center}	
\end{table}
%
%defined by the
%tag in the first decay as \mpp , \mnn , \mn or \mnb.
\subsection{T symmetry test}

\begin{figure}[htbp] %  figure placement: here, top, bottom, or page
   \centering
   \includegraphics[width=7cm]{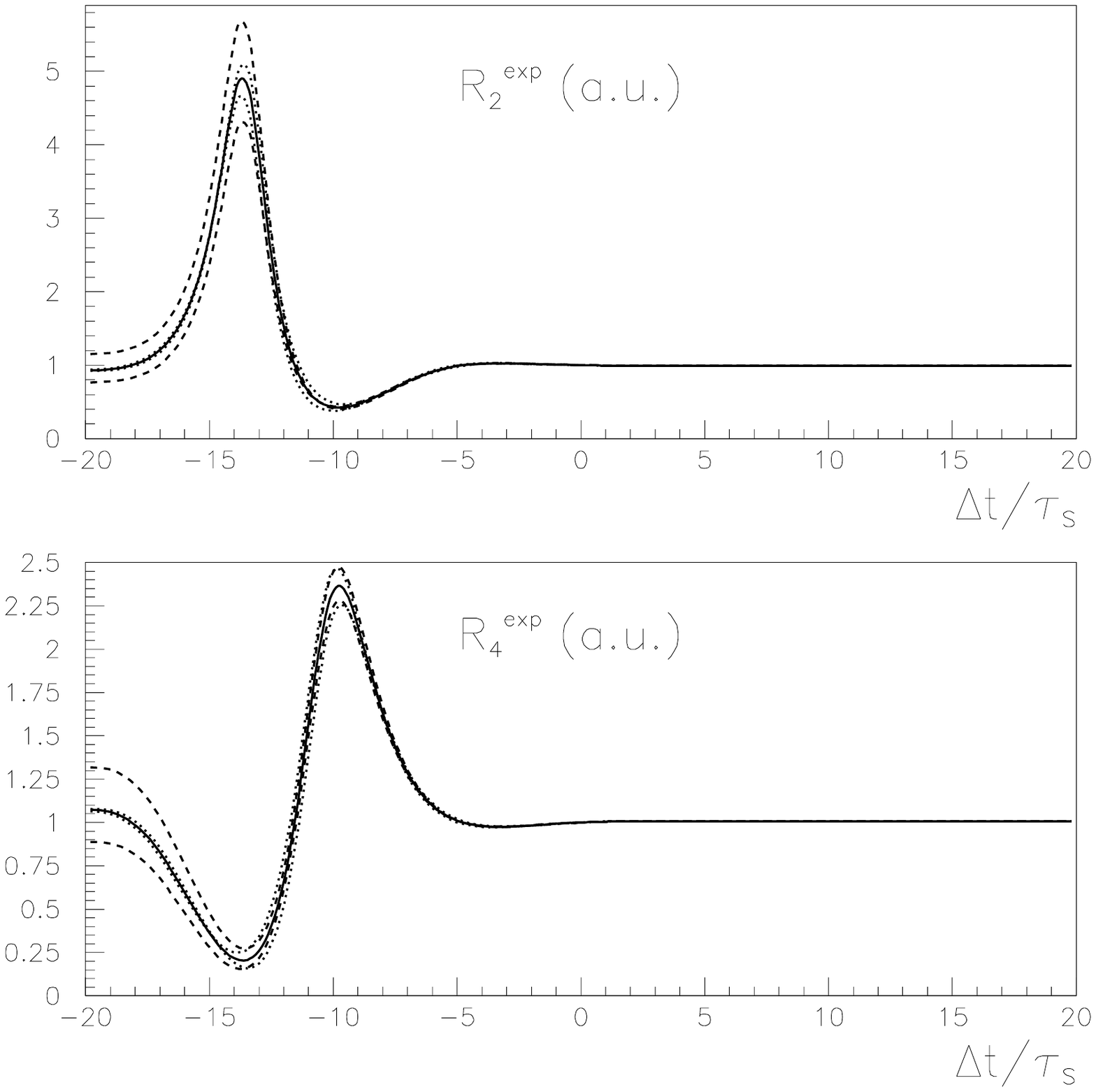} 
   \includegraphics[width=7cm]{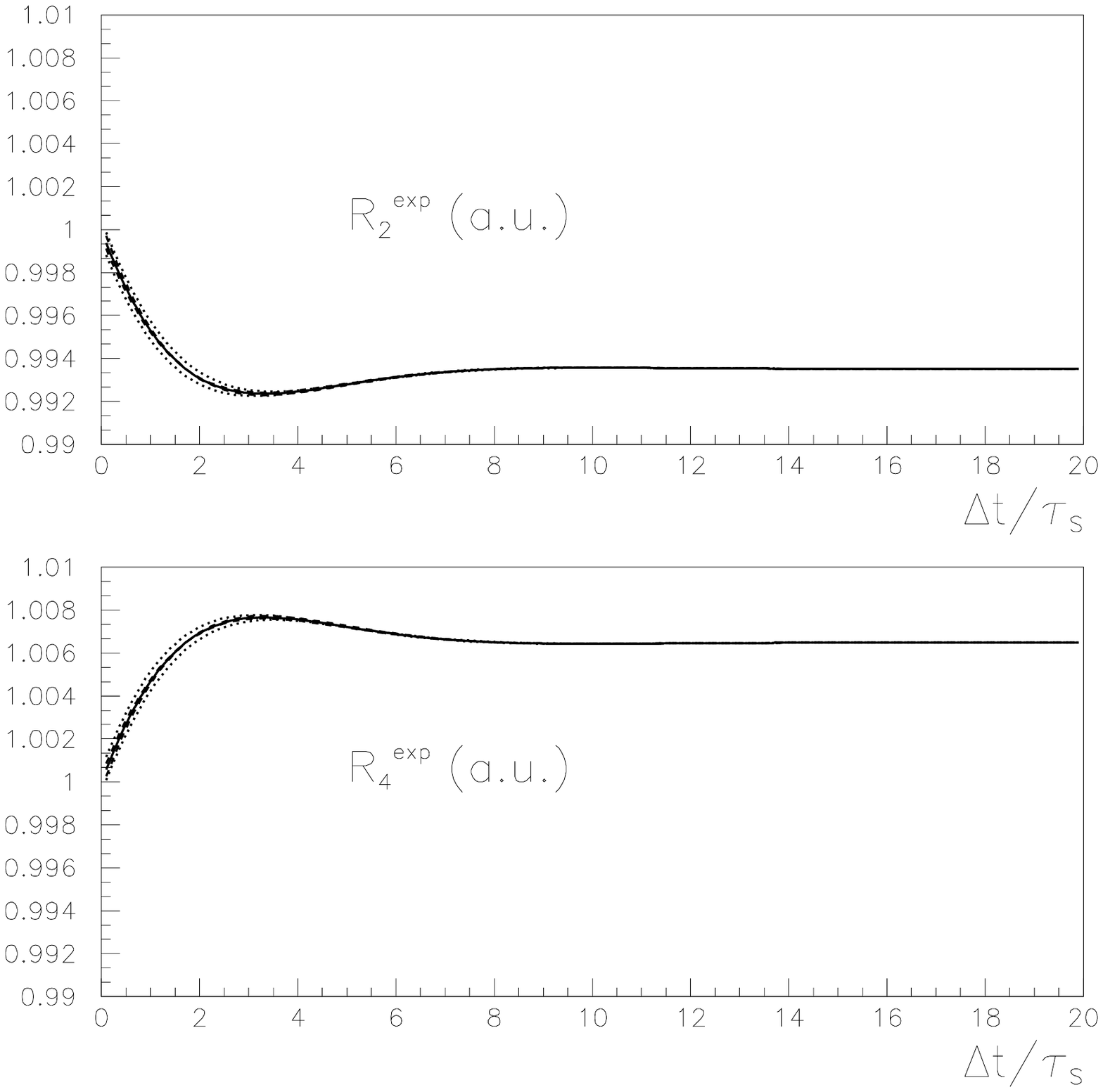} 
   \caption{The expected ratios $R_2^{\rm{exp}}(\Delta t)$ (left top) and $R_4^{\rm{exp}}(\Delta t)$ (left bottom)  as a function of 
   $\Delta t$ (solid line); dashed lines correspond to $\pm10\%$ uncertainty on $\eta_{3\pi^0}$; the constant $D$ has been fixed to one for simplicity. 
   Zoomed plots for $\Delta t>0$ (right).}
   \label{fig:fig3}
\end{figure}

For the direct T symmetry test one can define 
the following ratios of probabilities:
\begin{eqnarray}
R_1(\Delta t) &=& P\left[\kn(0)\to\kpp(\Delta t)\right]/P\left[\kpp(0)\to\kn(\Delta t)\right] \nonumber \\
R_2(\Delta t) &=& P\left[\kn(0)\to\knn(\Delta t)\right]/P\left[\knn(0)\to\kn(\Delta t)\right] \nonumber\\
R_3(\Delta t) &=& P\left[\knb(0)\to\kpp(\Delta t)\right]/P\left[\kpp(0)\to\knb(\Delta t)\right] \nonumber\\
R_4(\Delta t) &=& P\left[\knb(0)\to\knn(\Delta t)\right]/P\left[\knn(0)\to\knb(\Delta t)\right]~.
\label{eq:ratios}
%R_2 = P(\kn\to\knn)/P(\knn\to\kn) \\
%R_3 = P(\knb\to\kpp)/P(\kpp\to\knb) \\
%R_4 = P(\knb\to\knn)/P(\knn\to\knb)  
\end{eqnarray}
The measurement of any deviation from the prediction $R_i(\Delta t)=1$
%\begin{eqnarray}
%R_1(\Delta t)=R_2(\Delta t)=R_3(\Delta t)=R_4(\Delta t)=1
%\label{eq:tprediction}
%\end{eqnarray}
imposed by T invariance
%\par The measurement of any $R_i\neq 1$ 
is a signal of T violation. 
At a $\phi$-factory the corresponding observable quantities are two ratios,
$R_2^{\rm{exp}}(\Delta t)$ and $R_4^{\rm{exp}}(\Delta t)$ (see Fig.\ref{fig:fig3}), 
of double decay rates
%into specific decay products
%final states 
%$f_1$ 
%(e.g. for the kaon on the left side) 
%and $f_2$ 
%(e.g. for the kaon on the right side) 
as a function of the difference of kaon decay times $\Delta t$ \cite{ref:HandbookAD,tviol}; 
for $\Delta t >0$ one has (the first decay product symbol in parenthesis indicates the first in time of the two kaon decays):
\begin{eqnarray}
\label{eq:intensity2}
R_2^{\rm{exp}}(\Delta t) \equiv
\frac{  I(\ell^-,3\pi^0;\Delta t)}
{ I(\pi\pi,\ell^+;\Delta t)}   
= R_2(\Delta t \times D 
~~,~~
R_4^{\rm{exp}}(\Delta t) \equiv
\frac{  I(\ell^+,3\pi^0;\Delta t)}
{ I(\pi\pi,\ell^-;\Delta t)}   
= R_4(\Delta t) \times D
%= R_2(\Delta t)
%&\prop&
%P\left[ \knb(0)\to\knn(t_2-t_1) \right]
\end{eqnarray}  
while for $\Delta t <0$:
\begin{eqnarray}
\label{eq:intensity2}
R_2^{\rm{exp}}(\Delta t) 
%\equiv
%\frac{  I(\ell^-,3\pi^0;\Delta t)}
%{ I(\pi\pi,\ell^+;\Delta t)}   
=\frac{D}{R_3(|\Delta t|)}
~~,~~
R_4^{\rm{exp}}(\Delta t) 
%\equiv
%\frac{  I(\ell^+,3\pi^0;\Delta t)}
%{ I(\pi\pi,\ell^-;\Delta t)}   
=\frac{D}{R_1(|\Delta t|)}~.
%= R_2(\Delta t)
%&\prop&
%P\left[ \knb(0)\to\knn(t_2-t_1) \right]
\end{eqnarray}  
Here the normalization constant $D$, assuming no CPT violation in semileptonic decays, is $D=\{{\rm BR}\left( \kln\rightarrow 3\pi^0\right)  \cdot \Gamma_L \} 
/\{ {\rm BR}\left( \ksn\rightarrow \pi\pi\right) \cdot \Gamma_S \} $.\\

The KLOE-2 experiment at DA$\Phi$NE with an integrated luminosity of
$\mathcal{O}(10 \hbox{ fb}^{-1}  )$~\cite{kloe2epjc} could make a statistically significant T test,
measuring the ratios $R_2^{\rm{exp}}(\Delta t)$ and $R_4^{\rm{exp}}(\Delta t)$
%can largely improve the sensitivity considering a larger
integrated in the statistically most populated $\Delta t$ region,
%than $1~\tau_S$, e.g. the range 
$0\leq \Delta t \leq 300~\tau_S$ \cite{tviol}. Unfortunately in this region
$R_2^{\rm{exp}}(\Delta t)$ and $R_4^{\rm{exp}}(\Delta t)$ are expected to
be constant (see Fig.\ref{fig:fig3} right),
%in , 
and a precise knowledge of the normalization $D$ is needed in order to detect  T violation. This also implies that T violation in this $\Delta t$ region is constant, proportional to $\Re\epsilon$, i.e.
directly proportional to 
$\Delta\Gamma$, the width difference of the mass eigenstates. Therefore,
 in this $\Delta t$ region,
 T violation would not be present in the limit $\Delta\Gamma \to 0$,
 %contrary to what 
 as one might like
 % in a direct genuine test 
 \cite{ref:Wolfenstein}.

\begin{figure}[htbp] %  figure placement: here, top, bottom, or page
   \centering
   \includegraphics[width=7cm]{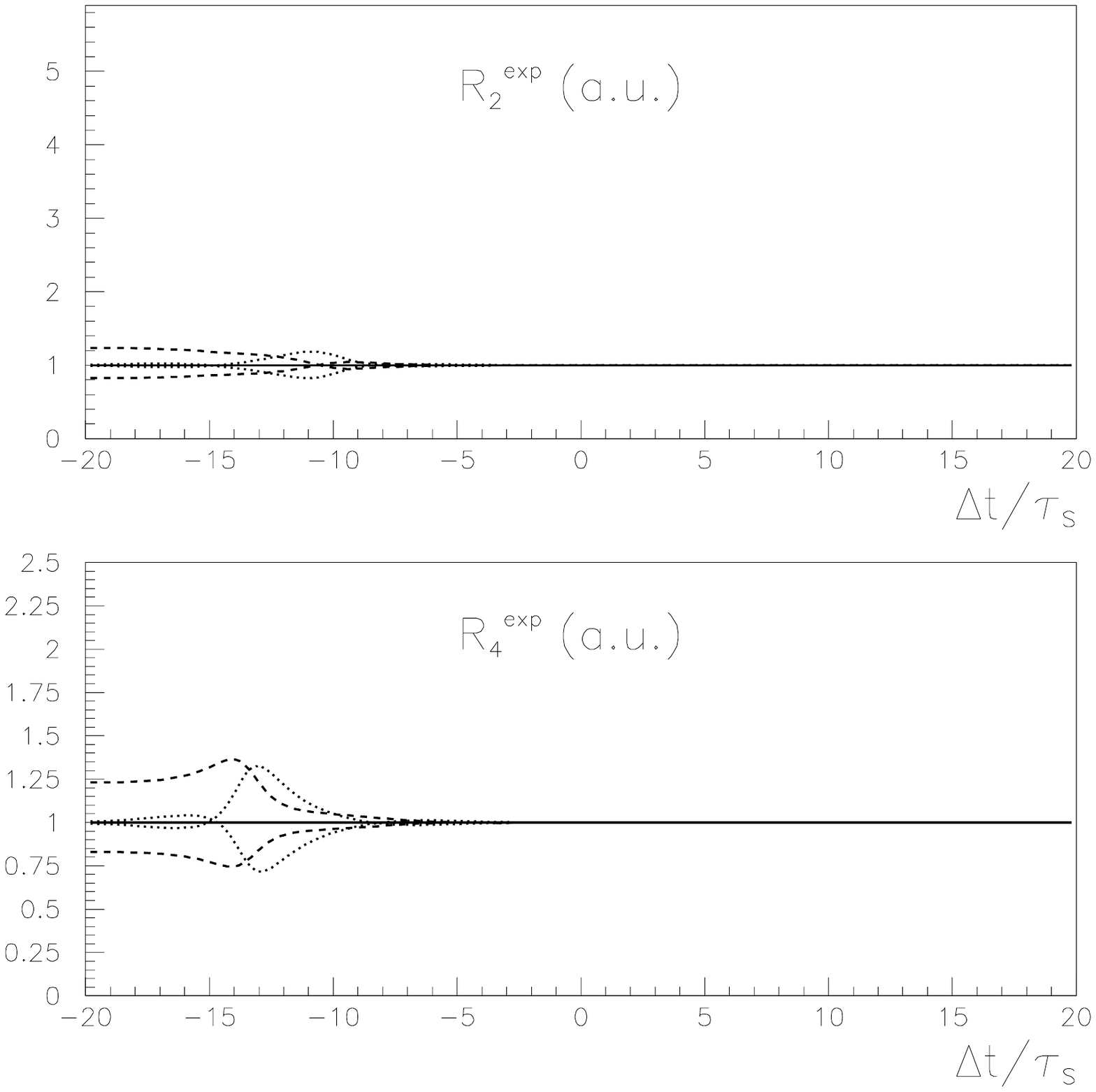} 
   \includegraphics[width=7cm]{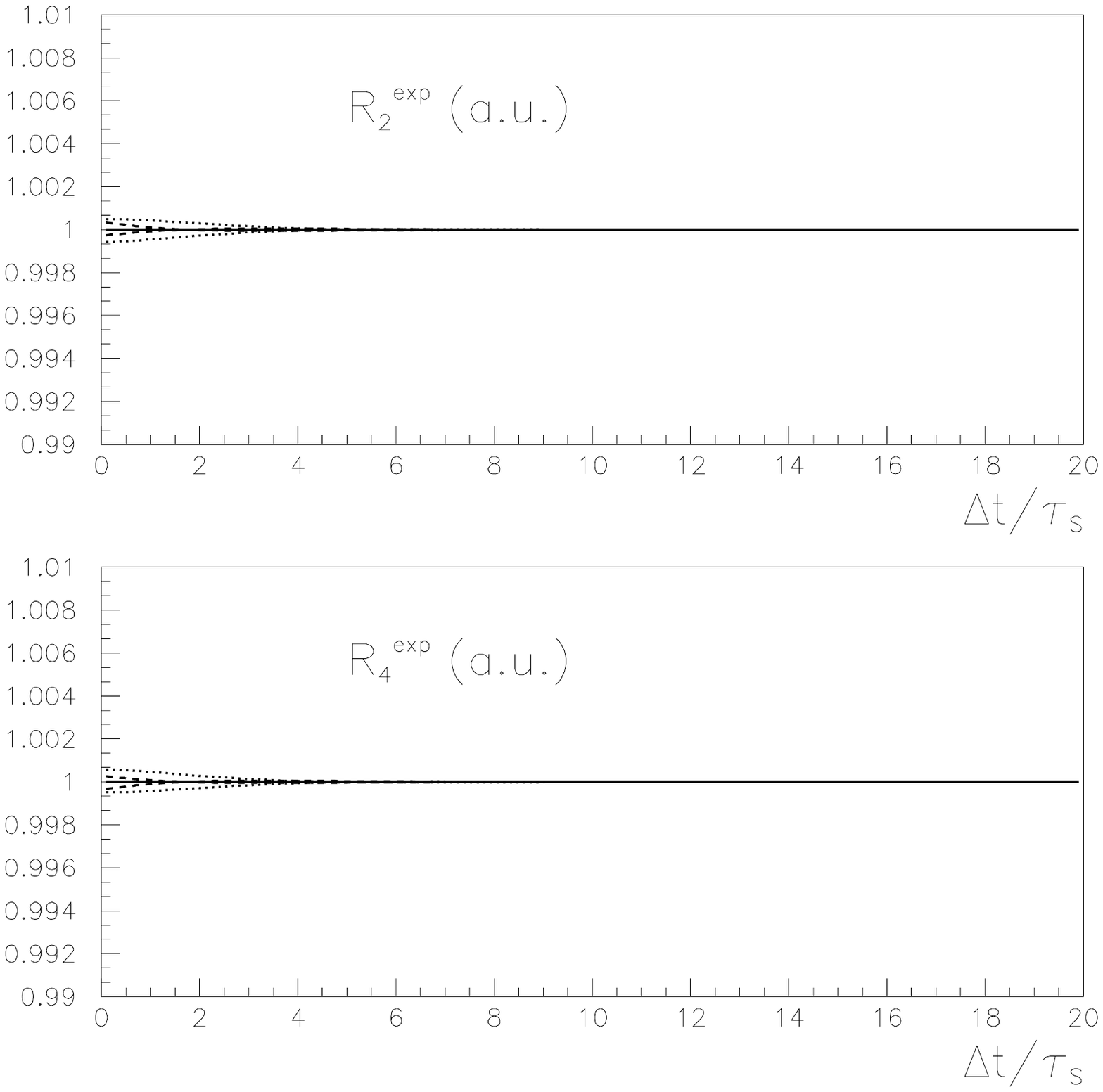} 
   \caption{The expected ratios $R_{2,CPT}^{\rm{exp}}(\Delta t)$ (left top) and $R_{4,CPT}^{\rm{exp}}(\Delta t)$ (left bottom)  as a function of $\Delta t$ (solid line); dashed lines correspond to $\pm 10\%$ uncertainty on $\eta_{3\pi^0}$; the constant $D$ has been fixed to one for simplicity. Zoomed plots for $\Delta t>0$ (right).}
   \label{fig:fig4}
\end{figure}

\subsection{CPT symmetry test}

For the direct CPT symmetry test one can define 
the following ratios of probabilities, similarly as for the \T test:
%\begin{linenomath*}
\begin{eqnarray}
R_{1,CPT}(\Delta t) &=& P\left[\kn(0)\to\kpp(\Delta t)\right]/P\left[\kpp(0)\to\knb(\Delta t)\right] \nonumber \\
R_{2,CPT}(\Delta t) &=& P\left[\kn(0)\to\knn(\Delta t)\right]/P\left[\knn(0)\to\knb(\Delta t)\right] \nonumber\\
R_{3,CPT}(\Delta t) &=& P\left[\knb(0)\to\kpp(\Delta t)\right]/P\left[\kpp(0)\to\kn(\Delta t)\right] \nonumber\\
R_{4,CPT}(\Delta t) &=& P\left[\knb(0)\to\knn(\Delta t)\right]/P\left[\knn(0)\to\kn(\Delta t)\right]~.
\label{eq:ratios}
%R_2 = P(\kn\to\knn)/P(\knn\to\kn) \\
%R_3 = P(\knb\to\kpp)/P(\kpp\to\knb) \\
%R_4 = P(\knb\to\knn)/P(\knn\to\knb)  
\end{eqnarray}
The measurement of any deviation from the prediction $R_{i,CPT}(\Delta t)=1$
%\begin{eqnarray}
%R_1(\Delta t)=R_2(\Delta t)=R_3(\Delta t)=R_4(\Delta t)=1
%\label{eq:tprediction}
%\end{eqnarray}
imposed by CPT invariance
%\par The measurement of any $R_i\neq 1$ 
is a signal of CPT violation. 
At a $\phi$-factory the corresponding observable quantities are, 
for $\Delta t >0$:
% one has (the first decay product symbol in parenthesis indicates the first of the two kaon decay):
\begin{eqnarray}
\label{eq:intensity2}
R_{2,CPT}^{\rm{exp}}(\Delta t) \equiv
\frac{  I(\ell^-,3\pi^0;\Delta t)}
{ I(\pi\pi,\ell^-;\Delta t)}   
= R_{2,CPT}(\Delta t) \times D_{CPT}\nonumber \\ 
%~,~~
R_{4,CPT}^{\rm{exp}}(\Delta t) \equiv
\frac{  I(\ell^+,3\pi^0;\Delta t)}
{ I(\pi\pi,\ell^+;\Delta t)}   
=R_{4,CPT}(\Delta t) \times D_{CPT}
%= R_2(\Delta t)
%&\prop&
%P\left[ \knb(0)\to\knn(t_2-t_1) \right]
\end{eqnarray}  
while for $\Delta t <0$:
\begin{eqnarray}
\label{eq:intensity2}
R_{2,CPT}^{\rm{exp}}(\Delta t) 
%\equiv
%\frac{  I(\ell^-,3\pi^0;\Delta t)}
%{ I(\pi\pi,\ell^+;\Delta t)}   
=\frac{D_{CPT}}{R_{1,CPT}(|\Delta t|)}
~~,~~
R_{4,CPT}^{\rm{exp}}(\Delta t) 
%\equiv
%\frac{  I(\ell^+,3\pi^0;\Delta t)}
%{ I(\pi\pi,\ell^-;\Delta t)}   
=\frac{D_{CPT}}{R_{3,CPT}(|\Delta t|)}~.
%= R_2(\Delta t)
%&\prop&
%P\left[ \knb(0)\to\knn(t_2-t_1) \right]
\end{eqnarray}  
Here the normalization constant $D_{CPT}$
is $D_{CPT}=\{{\rm BR}\left( \kln\rightarrow 3\pi^0\right)  \cdot \Gamma_L \} 
/
\{ {\rm BR}\left( \ksn\rightarrow \pi\pi\right) \cdot \Gamma_S \} $ without any assumption on CPT violation in semileptonic decays.\\

The KLOE-2 experiment  could make a statistically significant CPT test,
measuring the ratios $R_{2,CPT}^{\rm{exp}}(\Delta t)$ and $R_{4,CPT}^{\rm{exp}}(\Delta t)$
%can largely improve the sensitivity considering a larger
integrated in the statistically most populated $\Delta t$ region,
%than $1~\tau_S$, e.g. the range 
$0\leq \Delta t \leq 300~\tau_S$ \cite{tviol}.
In this region
$R_{2,CPT}^{\rm{exp}}(\Delta t)$ and $R_{4,CPT}^{\rm{exp}}(\Delta t)$ are expected to
be constant. Here CPT violation is
proportional to Re\,$\delta$, which do not vanish in the limit $\Delta\Gamma \to 0$,
escaping the previous controversy for the T test.
Most importantly the impact of direct \CP violation on the orthogonality condition
is completely negligible 
in this $\Delta t$ region (see Fig.\ref{fig:fig4} right), and does not affect the significance of the CPT test.
%which is more delicate than the T test.

\end{document}